\documentstyle[epsfig]{article}
= -2truecm

\begin{document}

\begin{center}
{\large\bf First order quantum phase transitions of the \\
XX spin-$1/2$ chain in a uniform transverse field} \vskip .6cm
{\normalsize Feng Pan,$^{a,~b}$ ~Nan Ma,$^{a}$~Xin Guan,$^{a}$ ~and
J. P. Draayer$^{b}$} \vskip .2cm {\small $^{a}$Department of
Physics, Liaoning Normal University, Dalian 116029, P. R.
China\vskip .1cm $^{b}$Department of Physics and Astronomy,
Louisiana State University, Baton Rouge, LA 70803-4001}
\end{center}

\begin{abstract}
Quantum phase transitional behavior of a finite periodic XX
spin-${1\over{2}}$ chain with nearest neighbor interaction in a
uniform transverse field is studied based on the simple exact
solutions. It is found that there are $[N/2]$ level-crossing points
in the ground state, where $N$ is the periodic number of the system
and $[x]$ stands for the integer part of $x$, when the interaction
strength and magnitude of the magnetic field satisfy certain
conditions. The quantum phase transitions are of the first order due
to the level-crossing. The ground state in the thermodynamic limit
will be divided into three distinguishable
quantum phases.
\end{abstract}

It is well known that the finite periodic XX spin-${1\over{2}}$
chain with nearest neighbor interaction in a uniform transverse
field is simply solvable. The result was first reported by Lieb et
al, and then by many others.$^{[2-6]}$ Similar models have been
attracted a lot of attention recently due to the fact that they may
be potentially helpful in quantum information processing$^{[7-9]}$
and realizable by using quantum dots, optical lattice, or spin
interaction systems.$^{[10-12]}$ These spin systems usually undergo
quantum phase transitions (QPTs) under certain conditions at zero
temperature, which can be characterized by non-analyticity in
properties of the ground state.$^{[13]}$ There are intimate links
between QPTs and entanglement in the systems.$^{[9, 14-16]}$ In this
Letter it will be shown that there are a series of level-crossing
points when the interaction strength and magnitude of the magnetic
field satisfy certain conditions, and the entanglement
measure$^{[17,18]}$ defined in terms of von Neumann entropy of
one-body reduced density matrix can be used to indicating both the
multi-particle entanglement and QPTs in the system.

The Hamiltonian of the model can be written as

$${\cal H}_{\rm XX}=-J\sum^{N}_{i=1}\left(S^{+}_{i}S^{-}_{i+1}+
S^{+}_{i+1}S^{-}_{i}\right)+h\sum^{N}_{i=1}S^{0}_{i},\eqno(1)$$
where $S^{\mu}_{i}$ ($\mu=+,-,0$) are spin operators satisfying the
SU(2) commutation relations:
$[S^{0}_{i},S^{\pm}_{j}]=\delta_{ij}S^{\pm}_{j}$,
$[S^{+}_{i},S^{-}_{j}]=2\delta_{ij}S^{0}_{j}$, $J>0$ is the nearest
neighbor interaction strength, $h$ is a uniform transverse field,
and the periodic condition $S^{\mu}_{i+N}=S^{\mu}_{i}$ is assumed.
These spin operators can be realized by the periodic-$N$ hard-core
boson operators with $S^{+}_{i}\rightarrow b^{\dagger}_{i}$,
$S^{-}_{i}\rightarrow b_{i}$, and $S^{0}_{i}\rightarrow
b^{\dagger}_{i}b_{i}-{1\over{2}}$, which satisfy $[b_{i},
b_{j}^{\dagger}]=\delta_{ij}(1-2b^{\dagger}_{j}b_{j})$,
$[b^{\dagger}_{i},b^{\dagger}_{j}]=[b_{i},b_{j}]=0$, and
$(b_{i})^{2}=(b^{\dagger}_{i})^2=0$. Thus, up to a constant, (1) can
be rewritten as

$$H_{\rm XX}=-{1-t\over{2}}\sum^{N}_{i=1}\left(b^{\dagger}_{i}b_{i+1}+
b^{+}_{i+1}b_{i}\right)+{t\over{2}}\sum^{N}_{i=1}b^{\dagger}_{i}b_{i},\eqno(2)$$
where, in order to investigate QPT behavior of the system, we have
set $J=(1-t)/2$ and $h=t/2$ with $0\leq t\leq 1$. Though the
neglected constant term in (2) is dependent on $t$, it only results
in a slight change in the positions of critical points, and the
phase transitional behavior of the system keeps unchanged. It is
clear that the ground state of the system is in the ferromagnetic
(unentangled) phase when $t=1$ and in the long-range order
(entangled) phase when $t=0$. Therefore, $t$ serves as the control
parameter of the system. By using the results shown in Refs. [1-6],
the $k$-`particle' wavefunctions of (2) can be expressed as

$$\vert k;(i_{1}i_{2}\cdots i_{k})\rangle=A^{\dagger}_{i_{1}}A^{\dagger}_{i_{2}}\cdots
A^{\dagger}_{i_{k}}\vert 0\rangle\eqno(3)$$ with $1\leq i_{1}\neq
i_{2}\neq\cdots\neq i_{k}\leq N$, where $\vert 0\rangle$ is the
boson vacuum and thus the SU(2) lowest weight state with
$S_{i}^{-}\vert 0\rangle=0~\forall~i$, and
$A^{\dagger}_{\mu}=\sum_{j=1}^{N}c_{j}^{(\mu)}b^{\dagger}_{j}$ with

$$c_{j}^{(\mu)}=\left\{
\begin{tabular}{c}
$e^{\imath 2\pi\mu j/N}~~~~~~~{\rm
for}~k={\rm odd}$,~~~~~~\\\\
$e^{\imath\pi(2\mu +1)j/N}~~{\rm for}~k={\rm
even}$~~~~\\
\end{tabular}\right.\eqno(4)$$
corresponding to the $\mu$-th set of eigenvectors of the matrix with
$\sum^{N-1}_{i=1}(E_{i i+1}+E_{i+1 i})-(-1)^{k}(E_{1 N}+E_{N 1})$,
in which $E_{ij}$ are the matrix units or generators of $U(N)$ in
the fundamental representation. The corresponding eigen-energy of
(3) is

$$E^{k}(t)=\sum_{\mu=1}^{k}\epsilon_{i_{\mu}}(t)~~{\rm with}~~
\epsilon_{i_{\mu}}(t)=\left\{
\begin{tabular}{c}
$\epsilon({\rm o}~,{i_{\mu}},t)=-(1-t)\cos{2\pi
i_{\mu}\over{N}}+t/2~~{\rm for}~k={\rm odd}$,~~~~~~\\\\
$\epsilon({\rm
e},~{i_{\mu}},t)=-(1-t)\cos{\pi(2i_{\mu}+1)\over{N}}+t/2~~{\rm
for}~k={\rm
even}$\\
\end{tabular}\right.\eqno(5)$$
with $N\geq 2$. Though the above results are analytic, it is still
not easy to write out those corresponding to a specific state
explicitly, especially to the ground state, from (3),(4), and (5)
directly. However, we have verified that the ground state energy for
periodic-$N$ chain is related to the following set of
eigen-energies:

$$E^{k}_{\min}(t)=\left\{
\begin{tabular}{c}
$\sum^{[k/2]}_{i=1}\epsilon({\rm
o},~i,t)+\sum^{[k/2]}_{i=0}\epsilon({\rm o},~N-i,t)~~{\rm for}~{\rm
odd}~k$,\\\\
$\sum^{[k/2-1]}_{i=1}\epsilon({\rm
e},~i,t)+\sum^{[k/2]}_{i=0}\epsilon({\rm e},~N-i,t)~~{\rm for}~{\rm
even}~k$\\
\end{tabular}\right.\eqno(6)
$$
with $k=0,1,\cdots,[N/2]$, where $[x]$ stands for the integer part
of $x$. It should be stated that the ground state energy at $t=1$
corresponds to $E^{0}_{\min}(t)=0$ from (6), while that at $t=0$
corresponds to $E^{[N/2]}_{\min}(t)$. Hence, it is clear that there
are $[N/2]+1$ different set of mutually orthogonal states with the
corresponding ground state energy $E^{[N/2]}_{\min}(t)$,
$E^{[N/2]-1}_{\min}(t)$, $\cdots$, $E^{1}_{\min}(t)$,
$E_{\min}^{0}(t)$, respectively, when the control parameter $t$
changes from $0$ to $1$. Such quantum phase transitions are of the
first order because the first derivative of the ground state energy
to the control parameter $t$ is discontinuous at the critical point,
$\lim_{t\rightarrow{t_{\rm c}-0}}{\partial E_{\rm
g}(t)\over{\partial t}}\neq \lim_{t\rightarrow{t_{\rm
c}+0}}{\partial E_{\rm g}(t)\over{\partial t}}$, according to the
extended Erhenfest classification of phase transitions.$^{[19]}$

\vskip .3cm \noindent{\bf Table 1.}~{First $9$ level-crossing points
for different $N$ cases
in addition to $t^{(0)}_{\rm c}$.}\\
\begin{tabular*}{\textwidth}{cccccccccc}
\hline \hline $N$~&$t^{(1)}_{\rm c}$& $t^{(2)}_{\rm c}$~
&$t^{(3)}_{\rm c}$ &$t^{(4)}_{\rm c}$& $t^{(5)}_{\rm c}$~
&$t^{(6)}_{\rm c}$ &$t^{(7)}_{\rm c}$
&$t^{(8)}_{\rm c}$ &$t^{(9)}_{\rm c}$\\
\hline\\
4&0.453082\\
6&0.594173&0.348915 \\
8&0.629014&0.531157&0.284603\\
10&0.643395&0.588789&0.478976&0.240565\\
12&0.650802&0.615444&0.551173&0.435657&0.208426\\
14&0.655138&0.630200&0.587316&0.517094&0.399305&0.18390\\
16&0.657902&0.639299&0.608381&0.560425&0.486483&0.36843&0.16456\\
18&0.659775&0.645332&0.621866&0.586711&0.535216&0.45901&0.34193&0.1489\\
20&0.661103&0.649550&0.631072&0.604041&0.565767&0.51177&0.43430&0.31895&0.13599\\
100&0.666447&0.666008&0.665347&0.664462&0.663352&0.66201&0.66043&0.65862&0.65657\\
1000&0.666664&0.66666&0.666654&0.666645&0.666634&0.66662&0.66660&0.66658&0.66656\\
 \hline \hline
\end{tabular*}
\vskip .3cm The first order phase transition in the system occurs
due to the ground state energy level-crossing of $E^{i}_{\min}(t)$
with $E^{i+1}_{\min}(t)$ for $i=0,1,\cdots,[N/2]-1$ with the
corresponding critical point $t^{(i)}_{\rm c}$, which is the root of
the simple linear equation $E^{i}_{\min}(t^{(i)}_{\rm
c})=E^{i+1}_{\min}(t^{(i)}_{\rm c})$ for $i=0,1,2,\cdots, [N/2]-1$.
Hence, there are $[N/2]$ critical points within the control
parameter range $0\leq t\leq 1$. Fig. 1 clearly shows the ground
state level crossings in the entire control parameter range for
$N=6,~8,~,20$ and $100$ cases. It is obvious that there are $[N/2]$
level-crossing points dividing the ground state into $[N/2]+1$
different parts, of which each is within a specific $t$ range when
$N$ is a finite number. With $N$ increasing, however, these specific
ranges become smaller and smaller, and finally tends to
infinitesimal, thus the ground state level becomes a continuous
phase before crossing to $E^{0}_{\min}$ level. Therefore, there will
be only one obvious critical point when $N\rightarrow\infty$. Since
$E^{1}_{\min}(t)=3t/2-1$ for any $N$, the obvious critical point is
at $t^{(0)}_{\rm c}=2/3$ in the thermodynamic limit. Nevertheless,
other level-crossing point $t^{(i)}_{\rm c}$ values in the finite
$N$ cases are $N$-dependent, of which some examples are listed in
Table 1.

\begin{center}
\begin{figure}[lh]
\center{\epsfig{file=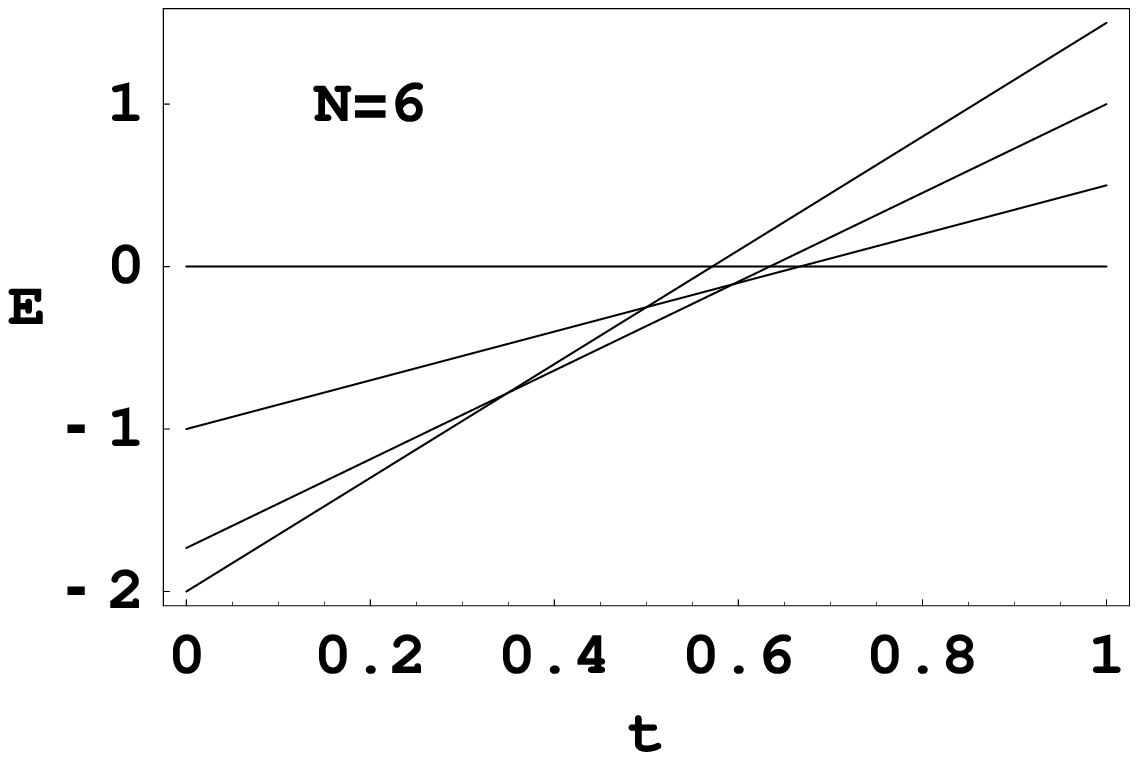,width=7.2cm}
\epsfig{file=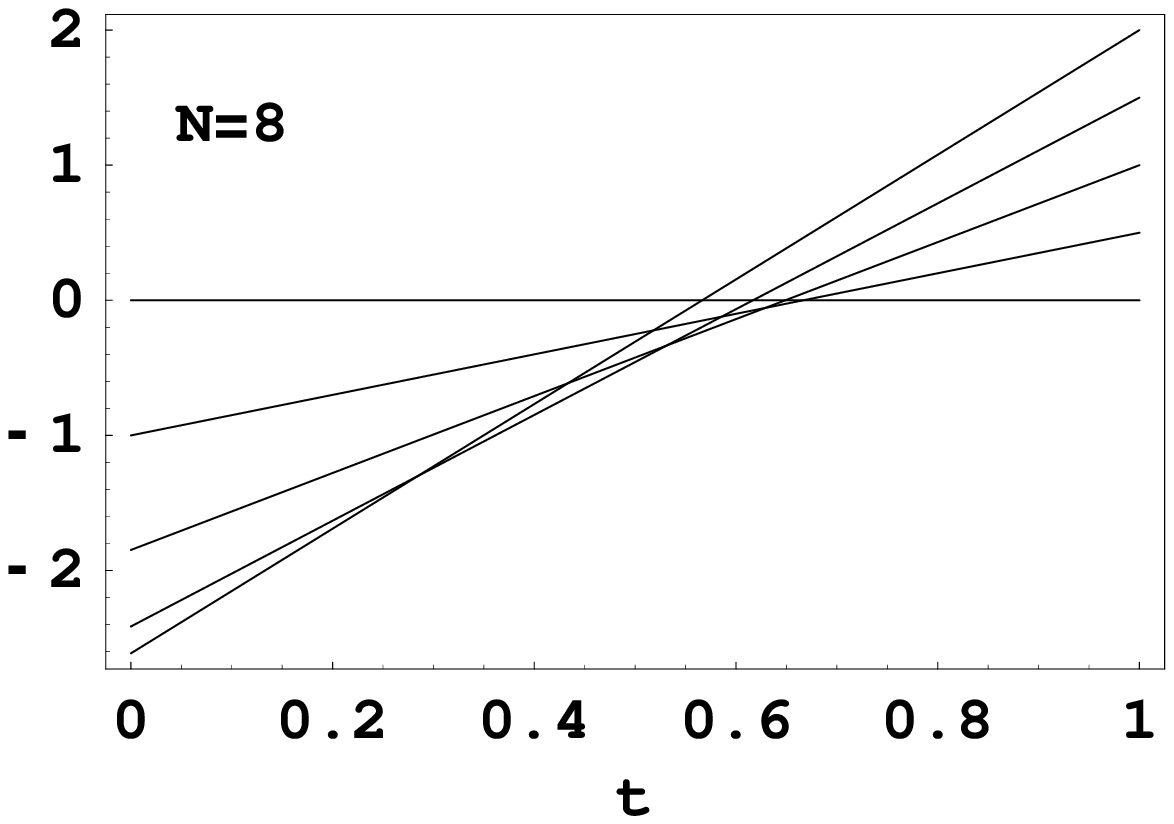,width=7.2cm}}\\
\center{\epsfig{file=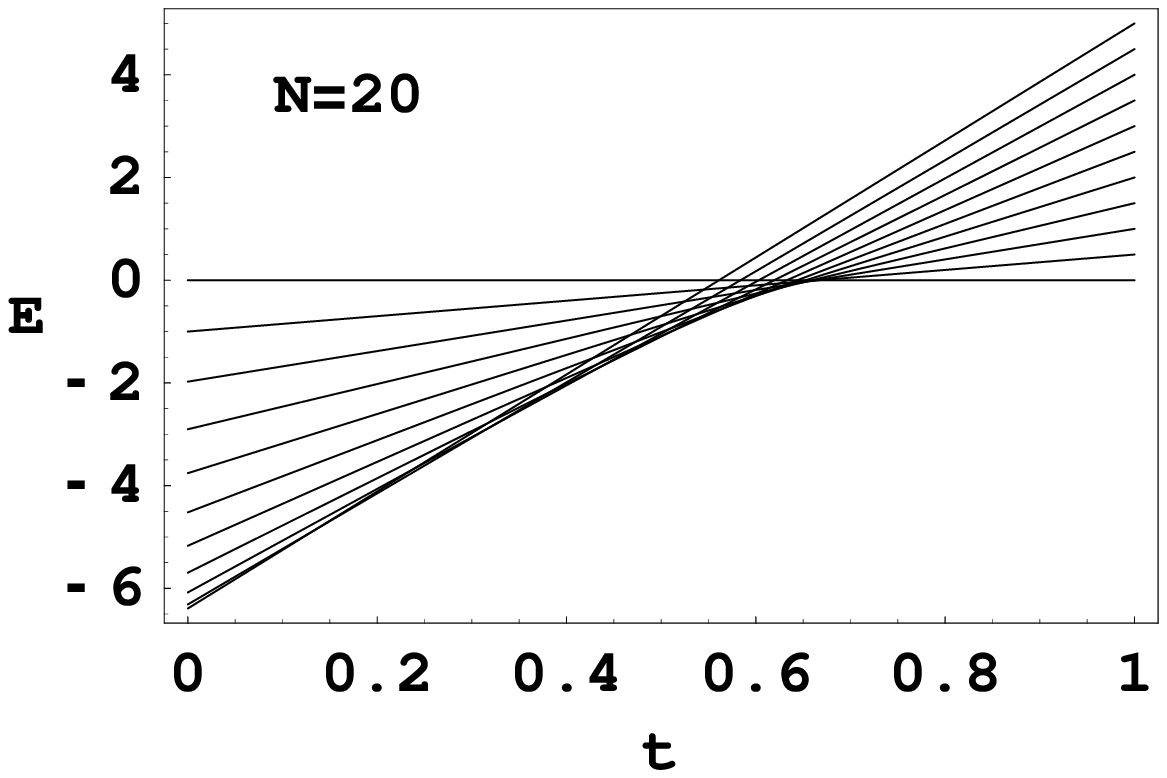,width=7.2cm}\epsfig{file=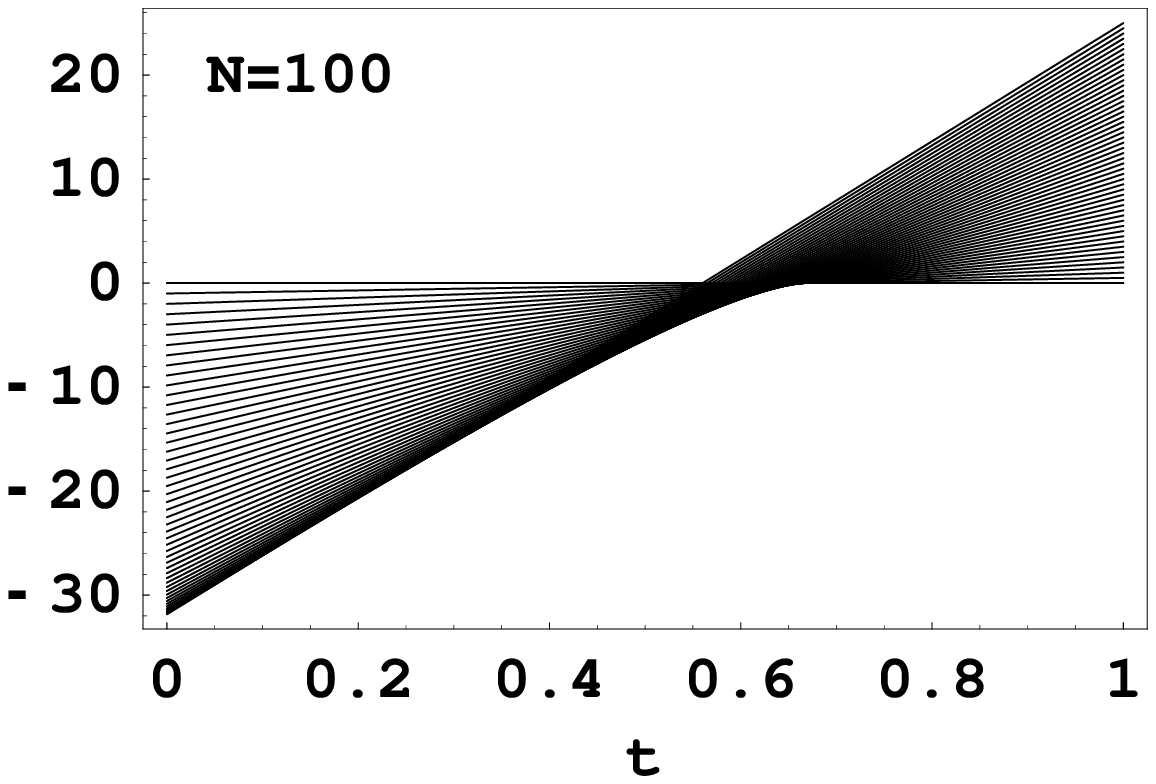,width=7.2cm
}} \caption{Level crossings related to the ground state for
different $N$ cases.}
\end{figure}
\end{center}

   Entanglement in the model is often studied by using
block--block entanglement defined in terms of von Neumann
entropy$^{[15]}$ or Wootters concurrence$^{[20]}$, e. g., that shown
in [21]. In the following, we use the simple measure proposed in
[17] with

$$\eta(\Psi)=
-{1\over{N}}\sum^{N}_{i=1}{\rm Tr}\left\{
(\rho_{\Psi})_{i}\log(\rho_{\Psi})_{i} \right\}\eqno(7)$$ if all $N$
terms in the sum are non-zero, otherwise $\eta(\Psi)=0$, where
$\Psi$ stands for the ground state wavefunction and
$(\rho_{\Psi})_{i}$ is the reduced density matrix with the $i$-th
site only. It has been shown$^{[18, 22,23]}$ that (7) is also
suitable to measure genuine $N$-body entanglement. We observed that
$(\rho_{\Psi})_{i}$ is independent of $i$ for the ground state in
the model. Hence, the entanglement measure $\eta$ can be simply
defined by the reduced von Neumann entropy for any site in this
case. Table 2 shows ground state entanglement in different $t$
ranges for $N=2,\cdots,6$, respectively, in which the entanglement
type of the ground state in each range is indicated. For example,
the state is a linear combination of several GHZ-like states for
$N=4$ with $0\leq t< 0.453082$, while it is a linear combination of
serval W-like states for $N=5$ with $0\leq t<0.552786$. It is clear
that the ground state entanglement measure gradually increases while
the control parameter $t$ decreases, which is also characterized by
the quantum number $S^{0}=\sum_{i}S^{0}_{i}$. In the separable
ferromagnetic phase, $S^{0}$ reaches its lowest value with
$S^{0}=-N/2$, while it becomes close to $0$ when $t<t_{c}^{[N/2]}$,
in which the spin-up and -down fermions are most strongly correlated
in comparison to that in other cases. In the most entangled
long-range order phase, even-$N$ systems are most entangled with
$\eta=1$ which is always greater than those of the nearest odd-$N$
systems. Furthermore, the ground state is not degenerate if the
control parameter $t$ does not at those $[N/2]$ level-crossing
points, while it becomes two-fold degenerate when $t=t_{c}^{(i)}$
for any $i$ as far as these states are concerned, which is mainly
due to the $S_{2}$ permutation symmetry defined by the permutation
of two sets of sites with $\{1,2,\cdots, [N/2]\}\rightleftharpoons
\{N-1,N-2,\cdots, N-[N/2]\}$. Nevertheless, these pairs of
degenerate states are still distinguishable from each other by the
quantum number $S^{0}$ with their difference $\Delta(S^{0})=\pm 1$
and by values of the entanglement measure of these two degenerate
states. As a consequence, the ground state in the thermodynamic
limit is not degenerate when $t=0$; it becomes two-fold degenerate
everywhere when the control parameter $t$ is within the half-open
interval $t\in (0, 2/3]$ because the level-crossing points are dense
everywhere in this control parameter range in the
$N\rightarrow\infty$ limit; and finally it becomes non-degenerate
again when $2/3<t\leq 1$. Therefore, the ground state should be
classified into three phases rather than two in the thermodynamic
limit. These three phases are one entangled GHZ-type phase at $t=0$
with $\eta=1$, one degenerate entangled W-type phase with $t\in
(0,2/3]$ and $0<\eta<1$, and one  non-degenerate fully separable
phase with $t\in (2/3,1]$ and $\eta=0$. It has been proved at least
for small $N$ cases that GHZ- and W-type states are inequivalent
under the SLOCC transformations.$^{[22-24]}$ We call the quantum
phase at $t=0$ hidden because the first derivative of the ground
state energy seems continuous at $t\in [0,\epsilon\rightarrow 0)$.

\vskip .4cm \noindent{\bf Table 2.}~{Ground state entanglement with
each quantum phase for $N=2,\cdots,6$}\\{\small
\begin{tabular*}{\textwidth}{lllll}
\hline \hline $N$~~~~~~~~~~~~~~~~~~~~~~~~~~~~~~~
~~~~~~~Entanglement type in each phase\\
\hline\\
2~~{\begin{tabular}{c}
Fully separable\\
($\eta=0$)~$2/3< t\leq 1$\\
\end{tabular}}
~{\begin{tabular}{c}
Bell~($\eta=1$)\\
 $0\leq t<2/3$\\
\end{tabular}}\\\\
3~~{\begin{tabular}{c}
The same as above\\
\end{tabular}}
~{\begin{tabular}{c}
W ~($\eta=0.918296$)\\
 $0\leq t<2/3$\\
\end{tabular}}\\\\
4~~{\begin{tabular}{c}
The same as above\\
\end{tabular}}
~{\begin{tabular}{c}
W ~($\eta=0.811278$)\\
 $0.453082<t<2/3$\\
\end{tabular}}
~{\begin{tabular}{c}
GHZ Combination\\
($\eta=1$)\\
 $0\leq t<0.453082$\\
\end{tabular}}\\\\
5~~{\begin{tabular}{c}
The same as above\\
\end{tabular}}
~{\begin{tabular}{c}
W ~($\eta=0.721928$)\\
 $0.552786<t<2/3$\\
\end{tabular}}
~{\begin{tabular}{c}
W  combination\\
($\eta=0.970951$)\\
 $0\leq t<0.552786$\\
\end{tabular}}\\\\
6~~{\begin{tabular}{c}
The same as above\\
\end{tabular}}
~{\begin{tabular}{c}
W ($\eta=0.650022$)\\
 $0.594173<t<2/3$\\
\end{tabular}}
~{\begin{tabular}{c}
W Combination\\
($\eta=0.918296$)\\
 $0.594173< t<0.348915$\\
\end{tabular}}
~{\begin{tabular}{c}
GHZ Combination\\
($\eta=1$)\\
 $0\leq t<0.348915$\\
\end{tabular}}\\\\
\hline \hline
\end{tabular*}}

\vskip .4cm In summary, the ground state of the finite periodic-$N$
XX spin-${1\over{2}}$ chain with nearest neighbor interaction in a
uniform transverse field is revisited by using the simple exact
solutions. The energy eigenvalues and the corresponding eigenstates
related to the ground state of the system are obtained analytically.
The results show how the ground state of the model evolves from the
ferromagnetic phase to the anti-ferromagnetic long-range order phase
with decreasing of the control parameter $t$ introduced. In
addition, we have shown that there are $[N/2]$ level-crossings in
the system, in which the middle part of long-range order phases will
become a continuous one in the large-$N$ limit leading to the
three-phase result in the thermodynamic limit. Such level-crossing
was also observed from a numerical study for specific $N$ cases of
XY spin chain,$^{[25]}$ and should be common in other spin
interaction systems in a uniform transverse field. Obviously, our
analytic and finite $N$ analysis provide with the microscopic
structure of the ground state of the model. Similar analysis for
other spin chain models may also be helpful, which will be discussed
elsewhere.

\vskip .5cm One of the authors (F. Pan) is grateful to Professor
Elliott Lieb for communications on the subject. Support from the
U.S. National Science Foundation (0500291), the Southeastern
Universities Research Association, the Natural Science Foundation of
China (10575047), and the LSU--LNNU joint research program (C192136)
is acknowledged.

\end{document}